\documentclass{amsart}
\usepackage{amssymb,amsthm}
\usepackage{geometry}                
\usepackage{graphicx}

\begin{document}

\author{Chisaki Nakaizumi}
\address[Chisaki Nakaizumi]{Life Science Center of TARA, University of Tsukuba, Japan}
\author{Toshie Matsui}
\address[Toshie Matsui]{Life Science Center of TARA, University of Tsukuba, Japan}
\author{Koichi Mori}
\address[Koichi Mori]{Research Institute of National Rehabilitation Center for Persons with Disabilities, Japan}
\author{Shoji Makino}
\address[Shoji Makino]{Life Science Center of TARA, University of Tsukuba, Japan}
\author{Tomasz M. Rutkowski$^*$}\thanks{$^*$Corresponding author}
\address[Tomasz M. Rutkowski]{Life Science Center of TARA, University of Tsukuba and
RIKEN Brain Science Institute, Japan}
\email[Corresponding author]{tomek@bci-lab.info}

\title{Head--related Impulse Response--based Spatial Auditory Brain--computer Interface$^\dag$}\thanks{$^\dag${\bf Author contributions:} Designed and performed the EEG experiments: CN, TMR. Analyzed the data: CN, TMR. Conceived the concept of the HRIR--based spatial auditory BCI paradigm: TMR, KM. Supported the project: SM, KM, TM. Wrote the paper: CN, TMR} 

\maketitle

\markleft{C. NAKAIZUMI ET AL.}

\begin{abstract}
This study provides a comprehensive test of the head--related impulse response (HRIR) to an auditory spatial speller brain--computer interface (BCI) paradigm, including a comparison with a conventional virtual headphone--based spatial auditory modality. Five BCI--naive users participated in an experiment based on five Japanese vowels. The auditory evoked potentials obtained produced encouragingly good and stable P300--responses in online BCI experiments. Our case study indicates that the auditory HRIR spatial sound paradigm reproduced with headphones could be a viable alternative to established multi--loudspeaker surround sound BCI--speller applications.
\end{abstract}

\tableofcontents

\section{Introduction}
\label{sec:introduction}

A brain-computer interface (BCI) is capable of providing a speller for disabled people with conditions such as amyotrophic lateral sclerosis (ALS). Although the currently successful visual modality may provide a fast BCI speller, patients at an advanced stage who are in a locked--in state cannot use the modality because they lose all intentional muscle control, including even blinking and movements of the eyes. An auditory BCI may be an alternative method because it does not require good eyesight. However, the modality is not as precise as the visual.
%

We propose an alternative method to extend the previously published spatial auditory BCI (saBCI) paradigm~\cite{MoonJeongBCImeeting2013} by making use of a HRIR for virtual sound image spatialization with headphone--based sound reproduction. 
Our research goal is a virtual spatial auditory BCI using HRIR--based spatialized cues in the part of the non--invasive, stimulus--driven, auditory modality which does not require long--term training. 
%
Experiments were conducted to reproduce and provide a comparison with previously reported vector--based amplitude panning (VBAP)--based spatial auditory experiments~\cite{MoonJeongBCImeeting2013}. The more precise HRIR--based spatial auditory BCI stimulus reproduction was used to simplify previously reported real sound sources generated with surround sound loudspeakers~\cite{iwpash2009tomek}. 
HRIR appends interaural intensity differences (IID), interaural time differences (ITD), and spectral modifications to create the spatial stimuli, while VBAP appends only IID. HRIR allows for more precise and fully spatial virtual sound image positioning, even without utilizing the user's owner HRIR measurements~\cite{book:auditoryNEUROSCIENCE}.

The next section of this paper describes the experiment set--up and the HRIR--based saBCI paradigm, together with EEG signal acquisition, pre--processing and classification steps. In the third section, the event related potentials (ERP), and especially the $P300$ response latencies are described, with a classification and discussion of the HRIR--based saBCI paradigm information transfer rate (ITR) results, including a comparison with the conventional method. Finally, the conclusions and future research directions are indicated.

\section{Methods}
\label{sec:methods}


All of the experiments were performed at the Life Science Center of TARA, University of Tsukuba, Japan. Five paid BCI-naive users participated in the experiments. The average age of the users was $21.6$ years (standard deviation $ 0.547$ years; five females).
The psychophysical and online EEG BCI experiments were conducted in accordance with \emph{The World Medical Association Declaration of Helsinki - Ethical Principles for Medical Research Involving Human Subjects}. The experiment procedures were approved and designed in agreement with the ethical committee guidelines of the Faculty of Engineering, Information and Systems at the University of Tsukuba, Japan.
Five Japanese vowels (a, i, u, e, o) were used in this experiment. The vowels were taken from a sound dataset of female voices~\cite{amano2009development}. The monaural sounds were spatialized using the public domain \textsc{CIPIC HRTF Database} provided by the University of California, Davis~\cite{cipicHRTF}. Each Japanese vowel was set on a horizontal plane at azimuth locations of $-80^{\circ}$,$-40^{\circ}$, $0^{\circ}$, $40^{\circ}$, $80^{\circ}$ for the vowels \emph{a, i, u, e, o}, respectively. 
The psychophysical experiments were conducted to investigate the response time and recognition accuracy. The users were instructed to respond by pressing the button as soon as possible after they perceived the \emph{target} stimulus, as in a classical oddball paradigm~\cite{bciBOOKwolpaw}. In a single experiment session, $20$~\emph{targets} and $80$~\emph{non-targets} were presented. 
An online EEG experiment was conducted to investigate the P300 response with BCI--naive users. The brain signals were collected with a biosignal amplifier system \textsf{g.USBamp} by g.tec Medical Engineering GmbH, Austria. The EEG signals were captured by sixteen active gel--based electrodes attached to the following head locations \emph{Cz, Pz, P3, P4, Cp5, Cp6, P1, P2, Poz, C1, C2, FC1, FC2,} and \emph{FCz}, as in the extended $10/10$ international system~\cite{Jurcak20071600}. The ground electrode was attached on the forehead at the \emph{FPz} location, and the reference  on the  user's left earlobe.
\textsc{BCI2000} software was used for the saBCI experiments to present stimuli and display online classification results. A single experiment was comprised of five sessions which contained $10$~\emph{target} and $40$~\emph{non-target} stimuli.  
The stimulus duration was set to $250$~ms, the interstimulus interval (ISI) to $150$~ms, and brain signals were averaged $10$~times for each vowel classification. 
The EEG sampling rate was set to $512$~Hz, and a $50$~Hz notch filter to remove electric power line interference was applied in a rejection band of $48-52$~Hz. The band pass filter was set with $0.1$~Hz and $60$~Hz cut-off frequencies. The acquired EEG brain signals were classified online by the \textsc{BCI2000} application using a stepwise linear discriminant analysis (SWLDA) classifier with features drawn from the $0\sim800$~ms ERP interval.

\section{Results and Discussion}
\label{sec:results}

\begin{figure}[t]
	\begin{minipage}{0.33\linewidth}
		\includegraphics[height=7.5cm]{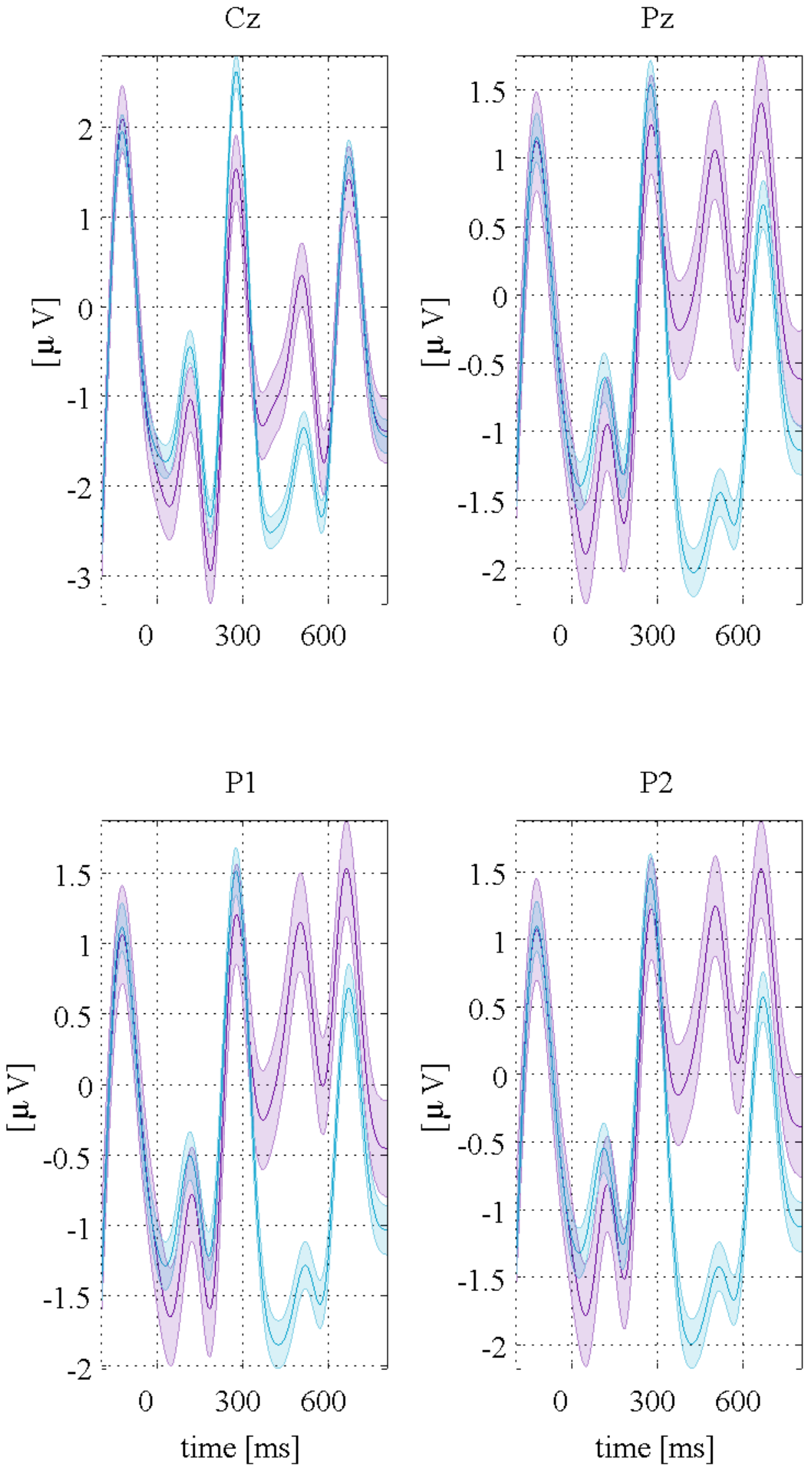}
	\end{minipage}
	 \begin{minipage}{0.6\linewidth}
	\includegraphics[height=7.5cm]{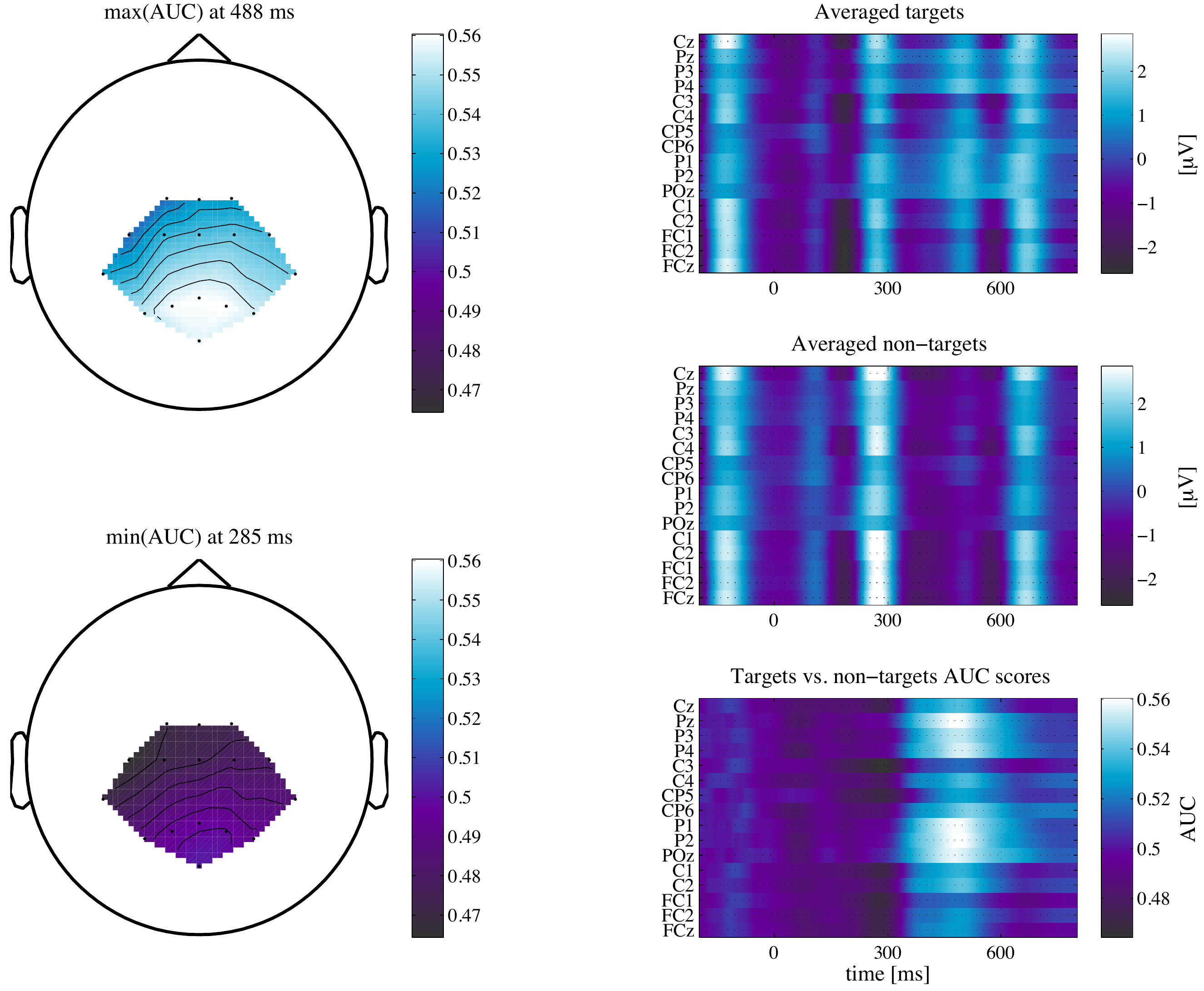}
	\end{minipage}
	\caption{Grand mean averaged ERP and AUC scores leading to final classification results for the participants. The left panel shows the averaged ERP for all participants. The purple line shows the brain waves for \emph{targets}, and blue line is for \emph{non-targets}. The centre panel presents the head topographies at the maximum and minimum AUC scores as obtained from the right bottom panel. The top right panel presents averaged ERP responses to the \emph{target} and the middle panel to the \emph{non--target} stimuli. The right bottom panel visualizes the AUC analysis results of \emph{target} versus \emph{non--target} response distribution differences.}\label{fig:EEGAUC2}
\end{figure}

This section presents and discusses results obtained from the psychophysical and EEG  experiments conducted with five users, as described in the previous section.
In the psychophysical experiment, the accuracy rates for all stimuli were above $94\%$. The majority of responses were concentrated at the $350$~ms latency. There were no significant differences in the response times between the target stimuli as tested by ANOVA $(p<0.05)$.
The results of the EEG experiment are depicted in Figure~\ref{fig:EEGAUC2}.
The left panel shows the grand mean averaged ERP results at four representative electrodes. The centre panel provides the results as scalp topographies at the maximum and minimum area under curve (AUC) of a receiver operating with characteristic values~\cite{bciSPATIALaudio2010} for \emph{target} vs. \emph{non--target} latencies. It also demonstrates the EEG electrode positions used in the experiments. The top right panel indicates the averaged ERP responses of all electrodes to the \emph{target}, and the second panel shows responses to the \emph{non--target} stimuli. The bottom panel indicates the AUC of \emph{target} versus \emph{non-target} responses, clearly confirming the usability of $400\sim600$~ms latencies for the subsequent classification.
Table~\ref{tab:EEG2} presents the classification accuracies of the P300 responses as obtained with the SWLDA classifier and the ITR scores. The average score was obtained as a mean value calculated from $1\sim5$~sessions (the training session was not included in the accuracies calculation). 
The ITR is a major comparison measure~\cite{bciSPATIALaudio2010} among the BCI paradigms. All five users scored above the five vowel sequences spelling chance levels of $20\%$. There was one user who achieved $100\%$ accuracy, which was the best in the experiments reported. 

We also compared the ITR scores with a VBAP--based spatial auditory BCI, which is regarded as a conventional method~\cite{MoonJeongBCImeeting2013}. The VBAP experiment was conducted in $2$~sessions and with $16$~BCI--naive users in \cite{MoonJeongBCImeeting2013}. The electrode positions were the same as in our current experiments. The sound stimuli were presented with small ear--fitting headphones in both the modalities. The ISI was set to $500$~ms in the VBAP experiment, and to $150$~ms in the HRIR experiment. 
In the VBAP modality, the average ITR score was $1.05$~bit/min and the best was $1.78$~bit/min. In the HRIR modality, the average ITR was $1.35$~bit/min and the best was $2.40$~bit/min. The ITR scores of the HRIR experiment were recalculated for $2$~sessions, the same as for the VBAP experiment. HRIR based modality produced better results than the VBAP based modality for both the average and the best score.

\section{Conclusions}
\label{sec:conclusions}

The EEG results presented confirm the P300 responses of BCI--naive users. The mean accuracy was not very good owing to the short ISI, but the accuracy tends to improve when the number of session increases. Therefore, more training may be necessary for BCI--naive users.

The ITR scores were higher compared with our previous study using HRIR stimuli, and also compared with the previously reported VBAP--based spatial auditory BCI.

Nevertheless, the current study is not able to compete with the faster visual BCI spellers. Furthermore, it is necessary to improve the ITR for a more comfortable spelling. We plan to continue research with larger numbers of sound stimuli, a better suited ISI, and more complex spatial sound patterns.
\begin{table}[tbp]
	\begin{centering}
	\caption{Vowel spelling accuracies and ITRs of each user obtained in the EEG experiments}\label{tab:EEG2}
	\begin{tabular}{ l | rrrrrrr | rr }
	\hline 
	\multicolumn{1}{c}{}
      & \multicolumn{7}{|c|}{Session}
      & \multicolumn{2}{c}{ITR [bit/min]}\\ \hline
User 	& 1~~~		& 2~~~		& 3~~~	& 4~~~ 		& 5~~~		& Average	& Best			& Average	& Best \\ 
\hline
$\#1$	& $60\%$  	& $80\%$  	&  $40\%$  &   $20\%$ &   $40\%$ 	& $48\%$	& $80\%$ 	& $2.26$ 	& $9.60$\\
$\#2$	& $0\%$    	& $40\%$  	&  $100\%$&   $80\%$ &   $100\%$	& $64\%$  	& $100\%$ 	& $5.27$ 	& $18.58$\\
$\#3$	& $20\%$  	& $20\%$  	&  $0\%$    &   $40\%$ &   $80\%$	& $32\%$	& $80\%$ 	& $0.46$ 	& $9.60$\\
$\#4$	& $20\%$  	& $20\%$  	&  $20\%$  &   $20\%$ &   $40\%$ 	& $24\%$	& $40\%$ 	& $0.06$ 	& $1.21$\\
$\#5$	& $0\%$    	& $40\%$  	&  $40\%$  &   $80\%$ &   $60\%$	& $44\%$	& $80\%$ 	& $1.70$	& $9.60$\\ \hline
	\end{tabular}
	\end{centering}
\end{table}

\section*{Acknowledgements}

The research was supported by the Strategic Information and Communications R\&D Promotion Program (SCOPE) no. 121803027 of The Ministry of Internal Affairs and Communications in Japan.

%


\newcommand{\etalchar}[1]{$^{#1}$}

\end{document}